\documentclass[aps,prl,twocolumn,amsmath,amssymb,nofootinbib,superscriptaddress,floatfix,reprint,longbibliography]{revtex4-1}
\usepackage[dvips]{graphicx}
\usepackage{latexsym}
\usepackage{amsmath}
\usepackage{amsfonts}
\usepackage{amssymb}
\usepackage{bm}
\usepackage{color}
\usepackage{txfonts}
\usepackage{float}
\usepackage{url}
\usepackage[version=4]{mhchem}
\usepackage{braket}
\usepackage{booktabs}
\usepackage{siunitx}
\usepackage{mhchem}
\usepackage[colorlinks=true, urlcolor=blue, linkcolor=blue, citecolor=blue, pdftex]{hyperref}
\usepackage{ulem}
\DeclareUnicodeCharacter{2212}{-}

\begin{document}
\title{Electronic structure and superconducting properties of LaNiO$_2$}

\author{Ziyan Chen}
\affiliation{Beijing National Laboratory for Condensed Matter Physics and Institute of Physics,
	Chinese Academy of Sciences, Beijing 100190, China}
\affiliation{School of Physical Sciences, University of Chinese Academy of Sciences, Beijing 100190, China}

\author{Yuxin Wang}
\affiliation{Beijing National Laboratory for Condensed Matter Physics and Institute of Physics,
	Chinese Academy of Sciences, Beijing 100190, China}
\affiliation{School of Physical Sciences, University of Chinese Academy of Sciences, Beijing 100190, China}

\author{Kun Jiang}
\email{jiangkun@iphy.ac.cn}
\affiliation{Beijing National Laboratory for Condensed Matter Physics and Institute of Physics,
	Chinese Academy of Sciences, Beijing 100190, China}
\affiliation{School of Physical Sciences, University of Chinese Academy of Sciences, Beijing 100190, China}

\author{Jiangping Hu}
\email{jphu@iphy.ac.cn}
\affiliation{Beijing National Laboratory for Condensed Matter Physics and Institute of Physics,
	Chinese Academy of Sciences, Beijing 100190, China}
\affiliation{Kavli Institute of Theoretical Sciences, University of Chinese Academy of Sciences,
	Beijing, 100190, China}
 \affiliation{New Cornerstone Science Laboratory, 
	Beijing, 100190, China}

\date{\today}

\begin{abstract}
Motivated by recent photoemission measurements on the La$_{0.8}$Sr$_{0.2}$NiO$_2$, we carry out a systematic study of the infinite-layer nickelate using both dynamical mean-field theory and density matrix embedding theory. The renormalized electronic structure and Fermi surface of correlated La$_{0.8}$Sr$_{0.2}$NiO$_2$ are studied in an effective two-band model through the dynamical mean-field calculation. We find the correlation effects reflect mainly on the Ni $d$ band, which is consistent with the experimental findings. We further study the ground state including magnetism and superconductivity through the density matrix embedding theory. Within the experimental doping range and rigid-band approximation, we show that the $d$-wave superconductivity is the lowest energy state, while the static magnetism is absent except very close to zero doping. These findings provide a new understanding of infinite-layer nickelate superconductivity.
\end{abstract}

\maketitle
The pairing mechanism of high-temperature (high $T_c$) superconductors (SCs) is one of the most important questions in condensed matter \cite{bednorz,doping_mott,keimer_review}.
Since the discovery of cuprates \cite{bednorz}, tremendous efforts have been spent on finding new high-temperature superconductors.
As a neighborhood of copper, nickelates have long been proposed as an ideal system for realizing high $T_c$ SC \cite{rice_PhysRevB.59.7901,pickett_PhysRevB.70.165109}.
This proposal becomes true after the infinite-layer nickelate Nd$_{0.8}$Sr$_{0.2}$NiO$_2$ (T$_c$ $\sim$ 15 K) thin films are synthesized using the soft-chemistry topotactic reduction method \cite{lidanfeng,norman}. 
This infinite-layer LnNiO$_2$ (Ln=La, Nd, Pr) opens a new window for nickelate SCs \cite{lidanfeng2,lidanfeng3,hwang,ariando22,Millis_PhysRevX.10.021061,Lechermann_PhysRevX.10.041002,Werner_PhysRevX.10.041047}. Recently, the family of nickelate SCs have been extended to La$_3$Ni$_2$O$_7$, La$_4$Ni$_3$O$_{10}$ under pressure \cite{meng_wang,yuanhq,chengjg,Kuroki_PhysRevB.109.144511,zhaojun,haihu_wen,yanpeng_qi}.

On the other hand, determining LnNiO$_2$ electronic structure is crucial to understanding its superconductivity.
Owing to their special structures and the CaH$_2$ reduction method \cite{lidanfeng}, it has been difficult to directly probe the electronic structure through angle-resolved photoemission spectroscopy (ARPES), scanning tunneling microscopy (STM) and other related techniques. 
Only very recently, the La$_{0.8}$Sr$_{0.2}$NiO$_2$ electronic structures were successfully measured by ARPES for the first time using the molecular beam epitaxy (MBE) and in situ atomic-hydrogen reduction methods \cite{yuefeng_nie,donglai_feng}.
These findings call for new theoretical calculations  to  determine correct minimal models for the nickelate superconductors.  
In this work, we carry out a systematic study of LnNiO$_2$ using  both modern dynamical mean-field theory (DMFT) and  density matrix embedding theory (DMET). We find that the  renormalized electronic structure and Fermi surface of correlated La$_{0.8}$Sr$_{0.2}$NiO$_2$  measured in the experiments can be quantitatively obtained  in an effective two-band minimal model through the dynamical mean-field calculation. Furthermore, we also predict that the $d$-wave superconductivity is the lowest energy state and there is no static magnetism  except close to zero doping for the nicklate superconductors.

\begin{figure}
    \centering
    \includegraphics[scale=0.25]{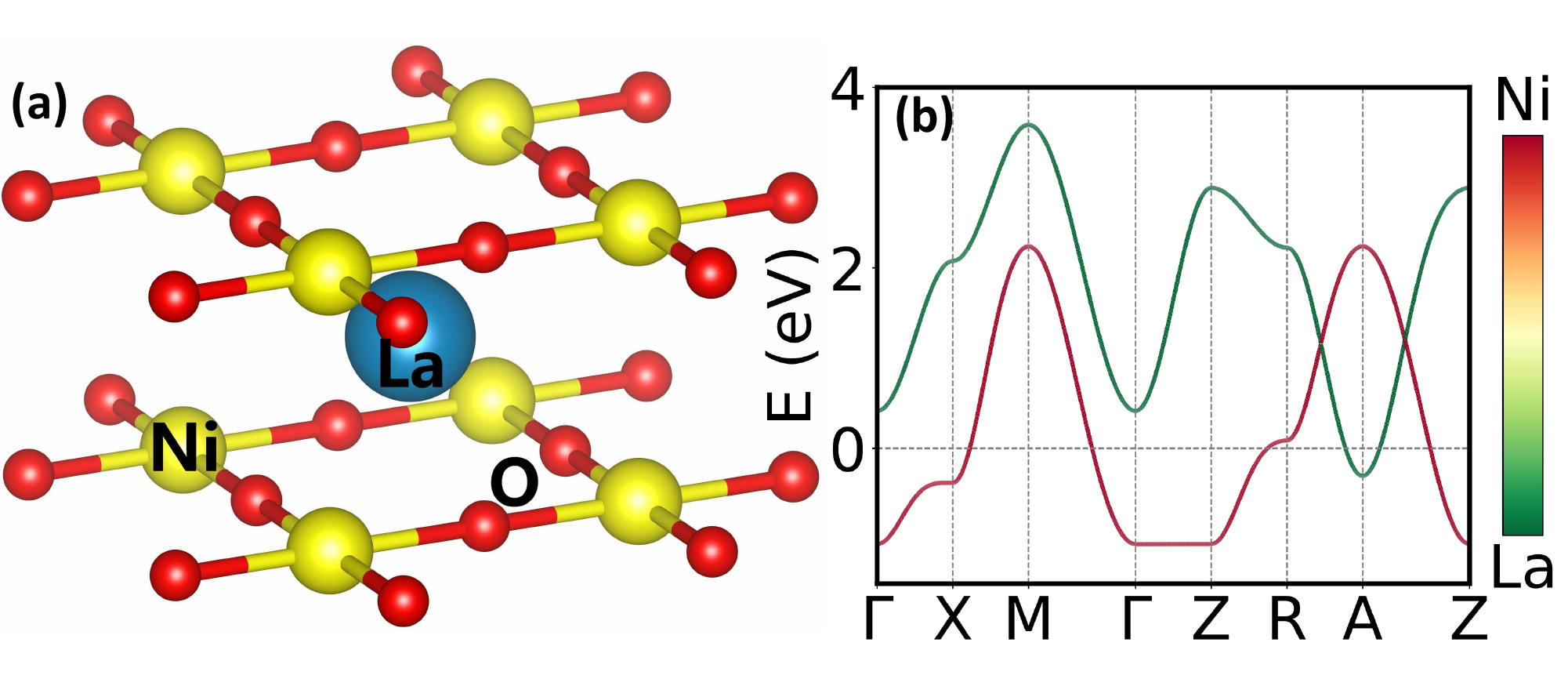}
    \caption{(a) Schematic plot of the LaNiO$_2$ crystal structure, where the blue, yellow, and red atoms represent La, Ni, and O elements, respectively; The central ingredient for LaNiO$_2$ electronic structure is the NiO$_2$ plane.
    (b) The band structure of the low-energy two-orbital effective model for La$_{0.8}$Sr$_{0.2}$NiO$_2$. The orbital content of Ni 3d$_{x^2-y^2}$ and La 5d are projected to the color bar between red (Ni) and green (La). 
    }
    \label{fig:structure_and_TB}
\end{figure}

Just like the cuprates SCs, the central component of LnNiO$_2$ is the NiO$_2$ plane, where Ni forms a square lattice and O sites at the bond connecting each Ni atom as shown in Fig.\ref{fig:structure_and_TB}(a). The Ln atoms are sandwiched between two NiO$_2$ planes.
Since the ARPES experiments focus on the La$_{0.8}$Sr$_{0.2}$NiO$_2$, we apply the density functional theory (DFT) to this hole-doped LnNiO$_2$. Our DFT calculations employ the Vienna ab-initio simulation package (VASP) code \cite{kresse1996efficient} with the projector augmented wave (PAW) method \cite{kresse1999ultrasoft}. The Perdew-Burke-Ernzerhof (PBE) \cite{perdew1996generalized} exchange-correlation functional is used in our calculation. The cutoff energy for expanding the wave functions into a plane-wave basis is set to be 500 eV. The energy convergence criterion is 10$^{-8}$ eV. The $\Gamma$-centered 13$\times$13$\times$15 k-meshes are used. The lattice relaxation parameters obtained from VASP calculations were $a = b = 3.93$ \AA, and $c = 3.38$ \AA. Then, based on the two-orbital model for infinite-layer nickelates proposed in previous studies \cite{weng_hongming}, we construct the low-energy effective two-band Hamiltonian to describe the electronic structure at hole doping $x=0.2$. The tight-binding (TB) model in the basis $(c_{1k\sigma},c_{2k\sigma})$ as
\begin{eqnarray}
		H_{t}(\textbf{k})=\left(\begin{array}{cc} 
		\epsilon_{11}(\textbf{k}) & \epsilon_{12}^*(\textbf{k}) \\
		\epsilon_{12}(\textbf{k}) & \epsilon_{22}(\textbf{k}) 
	\end{array}\right),\label{eq:tb}
\end{eqnarray}
where $1,2$ is the orbital index for Ni and La respectively and $\epsilon_{\alpha\beta}(\textbf{k})$ is the hopping function defined in the supplemental materials (SMs). The corresponding non-interacting electronic structure is plotted in Fig.\ref{fig:structure_and_TB}(b). We want to add a note here. The DFT calculated Fermi surface (FSs) is slightly different than the ARPES measurements \cite{yuefeng_nie,donglai_feng}. This is most likely owing to the changes in electronic structure caused by the correlation effect. We take the approximation that the correlation effect for La electrons is weak as in Ref. \cite{Millis_PhysRevX.10.021061}, which is absorbed into the TB parameters. Hence, we slightly change the TB parameters to match the experimental FSs. It turns out this is an efficient and effective choice. The comparison between TB and DFT is listed in SMs.

From the TB results in Fig.\ref{fig:structure_and_TB}(b), we can find that the Ni 3$d_{x^2-y^2}$ orbital (red color) contributes the most valence electrons around the Fermi level ($E_F$), which is similar to hole-doped cuprates with electron occupation between 3$d^9$ to 3$d^8$. 
On the other hand, most of La 5$d$ contributions are above $E_F$ with a small Fermi pocket around A point. 
This topological structure of FSs is consistent with experimental observation at this doping \cite{yuefeng_nie}. Notice that previous theoretical reports also show that there is another electron Fermi pocket around the $\Gamma$ point at this doping, which is absent in our results and experiments \cite{yuefeng_nie,donglai_feng}.
Additionally, owing to the $z$-direction coupling, the Ni 3$d_{x^2-y^2}$ electronic band is not strictly quasi-2-dimensional resulting in the van-hove point below $E_F$ (X point) at $k_z=0$ plane and above $E_F$ (R point) at $k_z=\pi$ plane.

After obtaining the TB model, we treat the electronic correlations beyond DFT to study band renormalization. Following previous strategy \cite{Millis_PhysRevX.10.021061} and experimental observations, 
we only add the Hubbard interaction to Ni 3d-electrons as
\begin{eqnarray}
    H_{int}=U \sum_{i} \hat{n}_{i1\uparrow}\hat{n}_{i1\downarrow}
\end{eqnarray}
and neglect the correlation effects of 5d-electrons as discussed above. 
Then, we employ the DMFT method to the $H_t+H_{int}$ and calculate the spectral function of La$_{0.8}$Sr$_{0.2}$NiO$_2$. The DMFT calculations are performed using the open-source TRIQS package \cite{Olivier_TRIQS,Alexander_solid_dmft} with the continuous-time Monte Carlo impurity solver based on the strong-coupling expansion \cite{Priyanka_TRIQSCTHYB}. The temperature is set to 290 K and $U$ is set to $3.1$ eV for the calculations as in Ref. \cite{Millis_PhysRevX.10.021061}.  The self-energy $\Sigma(\omega)$ is obtained by the analytical continuation from imaginary frequency to real frequency using the maximum entropy method \cite{MarkusAichhorn_MaximumEntropy}.


\begin{figure}
    \centering
    \includegraphics[scale=0.4]{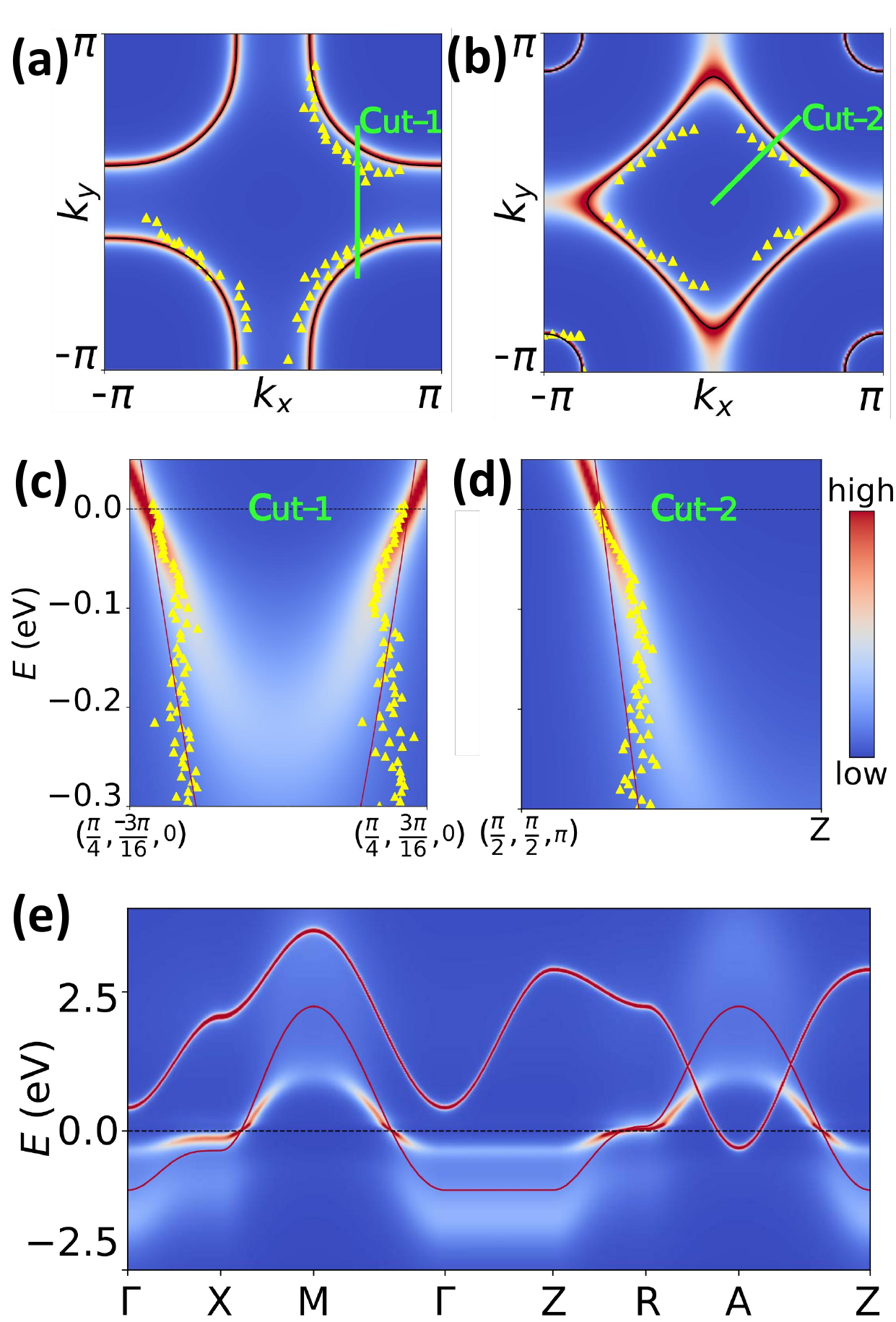}
    \caption{ (a-b) The Fermi surfaces and spectral function $A(k,\omega=0)$ plots at $k_z$=0 and $k_z$=$\pi$, respectively.
    (c), (d) The spectral function slices of La$_{0.8}$Sr$_{0.2}$NiO$_2$ obtained using the DMFT method along different $k$-paths.
    The spectral function slices of La$_{0.8}$Sr$_{0.2}$NiO$_2$ obtained using the DMFT method along Cut-1 and Cut-2 at (a-b). The yellow triangles represent the experimental data \cite{yuefeng_nie}, and the red solid line represents the TB results. (e) The renormalized $A(k,\omega)$ and non-interacting TB bands along the high-symmetry lines.}
    \label{fig:ARPS}
\end{figure}

We first look at the FSs and DMFT obtained spectral function $A(k,\omega=0)$ results as plotted in Fig.\ref{fig:ARPS} (a) and (b).
In the $k_z=0$ plane in Fig.\ref{fig:ARPS} (a), there is one FS centering around the Brillouin zone (BZ) corner, which shows a similar contour as hole-doped cuprates. The calculated FS matches well with the ARPES observed FS (yellow triangles) in Fig.\ref{fig:ARPS} (a) \cite{yuefeng_nie}. Then, moving to the $k_z=\pi$ plane, there are two groups of FSs. At the BZ corner, a small electron pocket from 5-d electron shows up as discussed above. 
At the BZ center, the 3d hole FS in $k_z=0$ plane evolves into the electron FS in $k_z=\pi$ plane. The feature indicates the Ni 3d-electron has a strong $k_z$ dependence, as discussed above. 
We also notice that the $A(k,\omega=0)$ in Fig. \ref{fig:ARPS} (a) remains sharp around the non-interacting Fermi points while the $A(k,\omega=0)$ in Fig. \ref{fig:ARPS} (b) becomes much broaden owing to the van-hove points. This distinct feature may influence the electronic properties of LnNiO$_2$.

Besides the FSs, the band renormalization effect can be captured well by the DMFT calculation. 
The renormalized spectral function $A(k,\omega)$ along the high-symmetry point is plotted in Fig.\ref{fig:ARPS} (e). 
We can see the 5d band remains unrenormalized owing to the uncorrelated approximation. As listed in SM, the 5d band mass from TB, DFT and experiments are almost the same without considering their bands' position shifting. 
This feature is consistent with the 5d band is highly overdoped from its half-filling resulting in a weakly renormalized band. On the other hand, the 3d quasiparticle $A(k,\omega)$ around $E_F$ indicates a renormalized band. In Fig. \ref{fig:ARPS}(c-d), the renormalized $A(k,\omega)$ and ARPES data are compared. We  find $A(k,\omega)$ agrees well  with experimental data around the Fermi level. In ARPES data, additional dispersion anomaly emerges below -$0.1$ eV, which is most likely a correlation effect beyond our current approach \cite{yuefeng_nie}. By comparing the Fermi velocity $v_F$ without renormalization and  $v_F$ at $A(k,\omega)$, we find the renormalization factor is about 0.3 at $k_z=0$ plane and 0.36 at $k_z=\pi$ plane.

We further explore the ground state of LaNiO$_2$. We apply the  DMET method developed in recent years \cite{Garnet_prl_DMET}.
The DMET is an efficient cluster embedding method for dealing with strongly correlated systems\cite{Garnet_prl_DMET}. Its basic principle is very similar to the DMFT method, where a lattice system is mapped onto a finite-size impurity problem. The density matrix between the bulk and impurity is matched after achieving self-consistency.
DMET has been applied the three-band model of cuprates \cite{cuizhihao_prr_3bandsHubbard,Garnet_jctc_DMET}. In the following calculations, we use the two-band model of La$_{0.8}$Sr$_{0.2}$NiO$_2$ discussed above. The Hubbard interaction $U = 3.1$ eV is applied to 3d-electrons while treating 5d-electrons as non-interacting.

\begin{figure}
    \centering
    \includegraphics[scale=0.46]{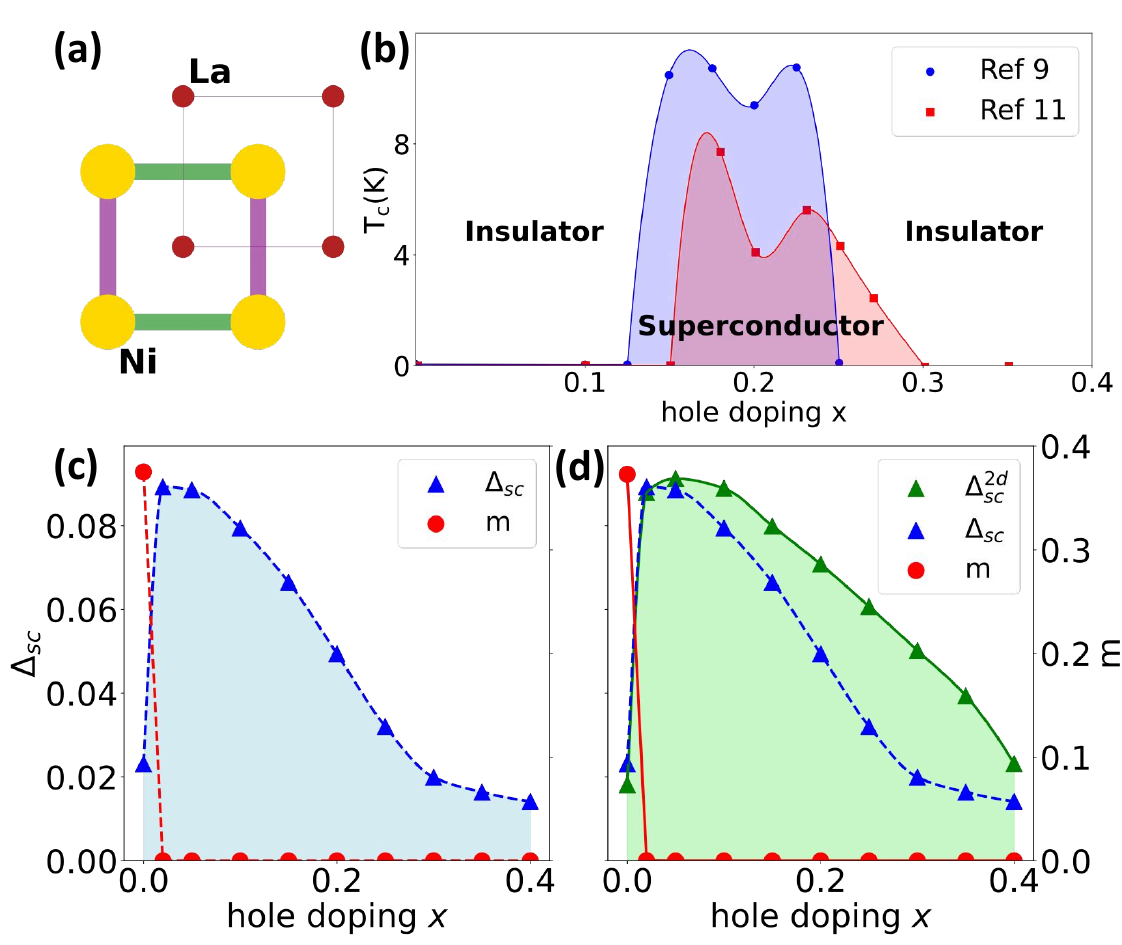}
    \caption{(a) The impurity cluster of LaNiO$_2$ for the DMET calculation. We use yellow and red to represent Ni and La atoms, respectively. The area of the circles represents the charge occupation on each atom, and the width and color of the lines between atoms represent the superconducting pairing strength and its sign.
    (b) The experimental phase diagrams of LnNiO$_2$ measured in Nd$_{1-x}$Sr$_x$NiO$_2$ \cite{lidanfeng3} and La$_{1-x}$Ca$_x$NiO$_2$ \cite{ariando22}.
    (c) The magnitude of the superconducting $\Delta_{SC}$ and magnetic order $m$ parameters at different levels of hole doping. The magnetic axis is at the right axis (red).
    The calculations were performed using a 2$\times$2 DMET cluster based on the TB model in Eq.\ref{eq:tb}, with the interaction on the Ni $d_{x^2-y^2}$ orbital set to $U = 3.1 \, \text{eV}$.
    (d) The phase diagram determined by DMET based on the TB model without z-direction hopping. The superconducting order parameter is labeled by $\Delta_{SC}^{2d}$ while the dash blue line represents the phase boundary in (c). }
    \label{fig:phase_diagram}
\end{figure}

In our  calculation, a 2$\times$2 supercell with 4 Ni and 4 La atoms is selected as the impurity cluster, as illustrated in Fig.\ref{fig:phase_diagram} (a). The total lattice size is chosen to be 40$\times$40$\times$20 unit cells.
The most relevant symmetry-breaking phases for the single-band Hubbard model are the antiferromagnetic (AFM) ordering at $(\pi,\pi)$ and $d$-wave superconductivity.
Hence, we introduce the pairing terms and magnetic terms in the correlation potential. The calculations focus on the experimental-related hole-doped region, up to a doping level $x$ of $0.4$. Furthermore, since the DMFT calculated FS of the La band is the same as the TB model, we fix the electron density of La as in the TB model at each self-consistent DMET calculation. The rigid band shift of the TB model is used for the doping change.

The DMET impurity solver used in our calculations is the open-source Block2 package \cite{HuanchenZhai_Block2}.
Block2 package implements the density matrix renormalization group (DMRG) as the impurity solver. The maximum bond dimension is chosen to be $800$. The tolerance for the DMRG sweep energy was set to $10^{-6}$. 
To extract the magnetic and superconducting properties of the system, we define the antiferromagnetic order parameter $m= \frac{1}{4} \sum_i \lvert m_i \rvert$,
where $m_i$ is the local magnetic moment on the Ni or La atoms, and the summation is taken over the four Ni atoms in the $2\times2$ supercell.
Similarly, the superconducting order parameter is defined as $\Delta_{SC} = \sum_{\langle ii' \rangle} \frac{1}{\sqrt{2}} \lvert  \langle d_{i\alpha} d_{i' \beta} \rangle + \langle d_{i' \alpha} d_{i \beta} \rangle  \rvert$
where the summation is taken over pairs of neighboring Ni or La atoms in the $2\times2$ supercell. The notation $\langle \dots \rangle$ refers to the reduced single-particle density matrix elements in the Nambu basis.

The self-consistent DMET calculations depend on the initial ansatz for the correlation potential, which determines the bulk symmetry-breaking effects. For the pairing part, we introduce $d$-wave superconducting pairing on the nearest-neighbor Ni orbitals, while pairing in other channels was not prohibited in the subsequent self-consistent calculations. For the magnetism part, we consider both paramagnetic and antiferromagnetic initial guesses. The convergence criteria are set to be less than $10^{-5}$ for the ground state energy and magnetic moments, and $10^{-4}$ for the superconducting order parameter. Typically, the calculations converge within 20 iterations.

The results of self-consistent DMET calculation based on the TB and Hubbard interaction are summarized in Fig.\ref{fig:phase_diagram}(c). Interestingly, no magnetism has been found in the hole-doping region from the DMET calculation except for $x$ very close to 0. 
There are two possible reasons for this feature. The Ni 3d-band is away from the half-filling in this region. At $x=0$, the filling fraction of 3d-electron is about $n_{Ni}=0.978$. It is known that the AFM of cuprates vanishes around $n_{Cu}=0.98\sim 0.95$ \cite{keimer_review} and the antiferromagnetic (AF) super-exchange coupling of Ni is weaker than Cu \cite{sawatzky_PhysRevLett.124.207004}. 
Furthermore, the itinerary La 5d-electron can further reduce the magnetism. Hence, the magnetic boundary is around $x=0$.

Considering  the superconducting order parameter $\Delta_{SC}$, a $d$-wave ground state is always obtained in this doping region. 
We also try the $s$-wave ansatz, which always evolves into the $d$-wave channel during iteration.
From the phase diagram in Fig.\ref{fig:phase_diagram}(c), the largest $\Delta_{SC}$ is obtained around $x=0.05$ which keeps decreasing with hole doping. The feature is consistent with previous cuprates calculations, where the electron-electron pairing is from AF spin exchange. The AF fluctuation becomes much stronger when close to its half-filling. 
The La 5d-electrons without correlation also acquire weak superconductivity owing to the coupling between Ni and La.

As discussed in the Fig.\ref{fig:ARPS}, the La$_{0.8}$Sr$_{0.2}$NiO$_2$ hosts a strong 3-dimension electronic structure.
This 3-dimensional feature may influence its superconducting order parameters. It is experimentally proposed that hydrogen can effectively tune the dimensionality of nickelates FSs \cite{qiaoliang-nature} inducing superconducting dome feature.
To capture this effect, we modify the hopping parameters in the TB model by ignoring the z-direction hopping. The resulting FSs become purely two-dimensional except for weak deviation owing to La electrons.
Then, we apply the same approach as above. The obtained phase diagram is shown in Fig.\ref{fig:phase_diagram} (d).
The magnetism is still missing although we have tried the AFM ansatz as the initial input. 
It is interesting that the dimensionality indeed influences the $\Delta_{SC}^{2d}$ order parameters, especially at the large doping region. This enhancement becomes much weaker when approaching $x=0$.
Hence, reducing the dimensionality of nickelates can indeed enhance the superconductivity.

Although DMET comprehensively studies the infinite layer nickelates, the phase diagram still deviates from the experiment findings \cite{lidanfeng3,ariando22}. 
Let's look at the up-to-date phase diagram of LnNiO$_2$ from various experimental groups \cite{lidanfeng3,ariando22}, as plotted in Fig.\ref{fig:phase_diagram} (b). There is a superconducting dome ranging from $x=0.1$ to $x=0.3$. Beyond this dome, two insulating regions emerge. LnNiO$_2$ is always a dirty and disordered system due to their reduction methods. Compared with cuprates, the disorder may play a more important role in LnNiO$_2$ physical properties.  

For the $x>0.3$ region, this region is already on the tail of our DMET SC region. It is well-known that the nodal $d$-wave superconductivity is more sensitive to disorder. Hence, no superconductivity identified in this region is reasonable due to reduction and doping-induced disorder.

On the contrary, the insulating $x<0.1$ region is completely different. In principle, the largest pairing order parameters in this region from our DMET calculation and previous understandings should be more robust than $0.1<x<0.3$ region.
Recently, a superconducting state has been reported near $x=0$ parent NdNiO$_2$ \cite{shen_nickelate,danfengli_conference}.
Therefore, we believe this region will still be superconducting if the sample is pure without disorder. The insulating behavior in this region is highly nontrivial owing to the interplay of disorder and strong correlation \cite{Hussey_insulating}, which calls for further exploration. Improving the sample quality and other controllable reduction methods is needed. 

In conclusion, we conduct a comprehensive study of infinite-layer nickelates using dynamical mean-field theory and density matrix embedding theory. Starting with density functional calculations and experimental data, we construct an effective two-band tight-binding model. We then use DMFT to examine the renormalized electronic structure and Fermi surface of the correlated compound La$_{0.8}$Sr$_{0.2}$NiO$_2$. We find that the correlation effects primarily impact the Ni 3d-band, in agreement with experimental observations. Leveraging this reliable model, we study the ground state properties of nickelates including magnetism and superconductivity via DMET. Within the doping range and under the rigid-band approximation, we identify $d$-wave superconductivity as the ground state and there is no static magnetism except close to $x=0$. The relation between our calculation and the experimental phase diagram is further discussed.
Our findings contribute to a new understanding of superconductivity in infinite-layer nickelates.

When finalizing this manuscript, we notice a similar DMFT study of nickelates \cite{held_dmft}, which arrives at a similar conclusion.

\textit{Acknowledgement} We thank Dr. Zhi-Hao Cui and Bo Zhan for the useful discussion and help in DMET. We acknowledge the support by the Ministry of Science and Technology  (Grant No. 2022YFA1403900), the National Natural Science Foundation of China (Grant No. NSFC-11888101, No. NSFC-12174428, No. NSFC-11920101005), the Strategic Priority Research Program of the Chinese Academy of Sciences (Grant No. XDB28000000, XDB33000000), the New Cornerstone Investigator Program, and the Chinese Academy of Sciences Project for Young Scientists in Basic Research (2022YSBR-048).

\bibliography{main.bbl}

\clearpage
\onecolumngrid
\begin{center}
\textbf{\large Supplemental Material}
\end{center}
\twocolumngrid

\setcounter{equation}{0}
\setcounter{figure}{0}
\setcounter{table}{0}
\setcounter{page}{1}
\makeatletter
\renewcommand{\theequation}{S\arabic{equation}}
\renewcommand{\thefigure}{S\arabic{figure}}
\renewcommand{\thetable}{S\arabic{table}}

\section*{APPENDIX A: Band structure}
The two bands Hamiltonian for LnNiO$_2$ can be written in the following form:
\begin{equation}
H(k) = \begin{pmatrix}
    H_{11} & H_{12} \\
    H_{21} & H_{22}
\end{pmatrix}
\end{equation}
where subscripts 1 and 2 correspond to the Ni and La atoms, respectively. The specific form of the diagonal terms is given by:
\begin{align}
H_{\alpha \alpha} &= t^{(0,0,0)}_{\alpha \alpha} + 2t^{(1,0,0)}_{\alpha \alpha} \left(\cos k_x + \cos k_y\right ) \notag \\
&\quad + 2t^{(0,0,1)}_{\alpha \alpha} \cos k_z + 4t^{(1,1,0)}_{\alpha \alpha} \cos k_x\cos k_y \notag \\
&\quad + 4t^{(1,0,1)}_{\alpha \alpha} \cos k_z\left(\cos k_x + \cos k_y \right) \notag \\
&\quad + 8t^{(1,1,1)}_{\alpha \alpha} \cos k_x\cos k_y\cos k_z \notag \\
&\quad +2t^{(1,0,0)}_{\alpha \alpha} \left(\cos2k_x + \cos2k_x\right)
\end{align}
The values of the hopping matrix elements are listed in Table.\ref{table_3D}. Here, \(H_{12}\) denotes the hopping between the La and Ni atoms. Due to symmetry considerations, the nearest-neighbor hopping matrix element vanishes, and the next-nearest-neighbor hopping takes the following form:
\begin{equation}
H_{12} = t^{(1,0,0)}_{12} (1 + e^{ik_z}) \left(2 \cos k_x - 2 \cos k_y \right)
\end{equation}
Naturally, we have \(H_{21} = H_{12}^*\).
\begin{table}
    \centering
    \caption{The hopping parameters for the two-band model Hamiltonians(with z-direction hopping).All the units are eV.}
    \setlength{\tabcolsep}{16pt}
    \label{table_3D}
    \begin{tabular}{c c|c c}
        \hline\hline
        $t_{11}^{(0,0,0)}$ & 0.4024 & $t_{22}^{(0,0,0)}$ & 1.8872 \\ \hline
        $t_{11}^{(1,0,0)}$ & -0.4125 & $t_{22}^{(1,0,0)}$ & 0.0013 \\ \hline
        $t_{11}^{(0,0,1)}$ & -0.0538 & $t_{22}^{(0,0,1)}$ & 0.0650 \\ \hline
        $t_{11}^{(1,1,0)}$ & 0.0894 & $t_{22}^{(1,1,0)}$ & -0.0606 \\ \hline
        $t_{11}^{(1,0,1)}$ & 0.000 & $t_{22}^{(1,0,1)}$ & -0.1980 \\ \hline
        $t_{11}^{(1,1,1)}$ & 0.0134 & $t_{22}^{(1,1,1)}$ & 0.0281 \\ \hline
        $t_{11}^{(2,0,0)}$ & -0.0430 & $t_{22}^{(2,0,0)}$ & 0.000 \\ \hline
        $t_{12}^{(1,0,0)}$ & 0.0196 & & \\ \hline\hline
    \end{tabular}
\end{table}

In our calculations, we also investigated the effect of the system's two-dimensionality on its superconducting properties. Specifically, we restricted the hopping of electrons in the Ni \(d_{x^2-y^2}\) orbitals to within the X-Y plane by manually setting the hopping matrix elements in the Z direction to zero. The Hamiltonian retains the same form as previously given, with parameters as listed in Table \ref{table_2D}.

\begin{figure}
    \centering
    \includegraphics[scale=0.45]{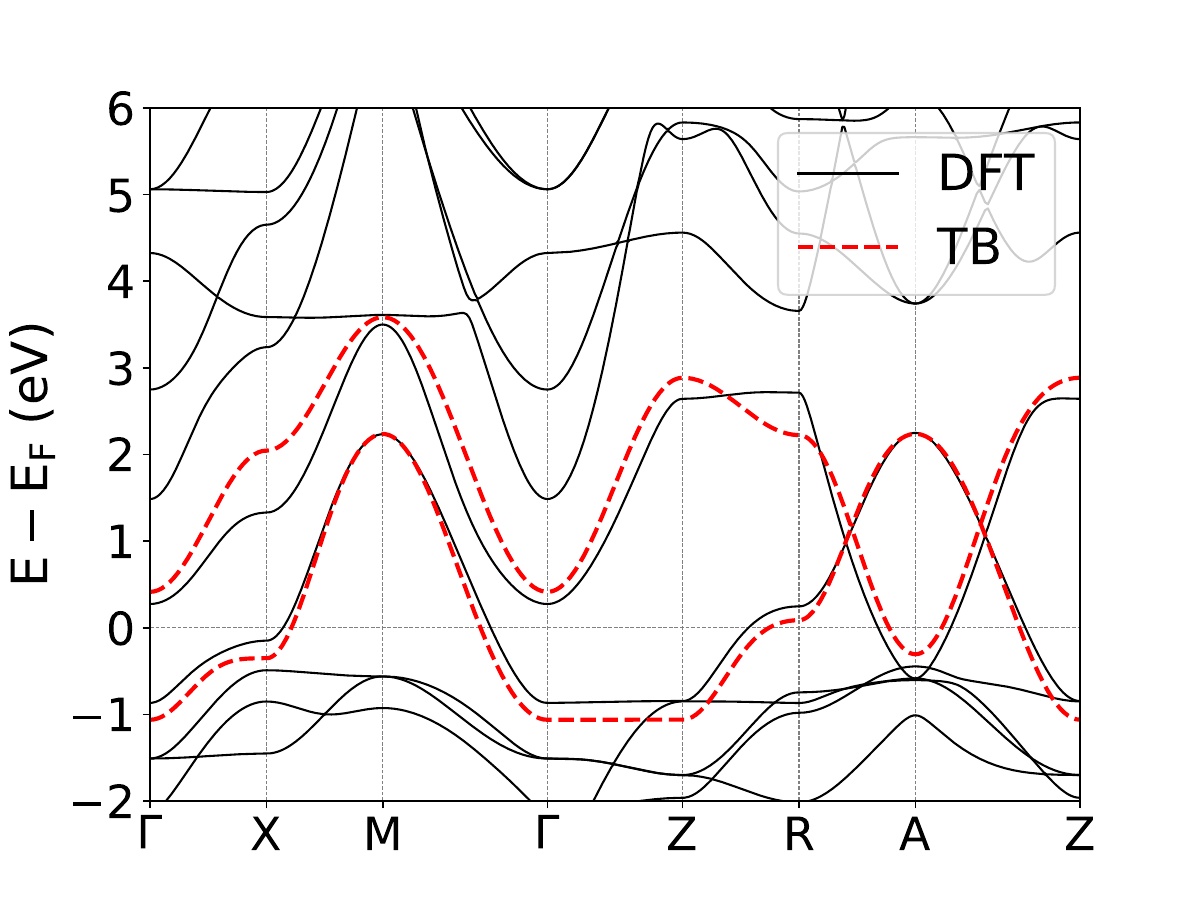}
    \caption{A comparison between Density Functional Theory (DFT) and the two-band model along high-symmetry points in the reciprocal space. The black solid lines represent the DFT results, while the red dashed lines correspond to the two-band model used in our work.}
    \label{fig:DFT and TB}
\end{figure}

In Fig.\ref{fig:DFT and TB}, the blue lines represent the band structure results obtained from Density Functional Theory (DFT) calculations, while the black dashed lines correspond to the band structure derived from the Hamiltonian using the parameters listed in Table \ref{table_3D}.

\begin{table}
    \centering
    \caption{The hopping parameters for the two-band model Hamiltonians(without z-direction hopping).All the units are eV.}
    \setlength{\tabcolsep}{16pt}
    \label{table_2D}
    \begin{tabular}{c c|c c}
        \hline\hline
        $t_{11}^{(0,0,0)}$ & 0.4024 & $t_{22}^{(0,0,0)}$ & 1.8872 \\ \hline
        $t_{11}^{(1,0,0)}$ & -0.4125 & $t_{22}^{(1,0,0)}$ & 0.0013 \\ \hline
        $t_{11}^{(0,0,1)}$ & 0.000 & $t_{22}^{(0,0,1)}$ & 0.0650 \\ \hline
        $t_{11}^{(1,1,0)}$ & 0.0894 & $t_{22}^{(1,1,0)}$ & -0.0606 \\ \hline
        $t_{11}^{(1,0,1)}$ & 0.000 & $t_{22}^{(1,0,1)}$ & -0.1980 \\ \hline
        $t_{11}^{(1,1,1)}$ & 0.000 & $t_{22}^{(1,1,1)}$ & 0.0281 \\ \hline
        $t_{11}^{(2,0,0)}$ & -0.0430 & $t_{22}^{(2,0,0)}$ & 0.000 \\ \hline
        $t_{12}^{(1,0,0)}$ & 0.0196 & & \\ \hline\hline
    \end{tabular}
\end{table}

\section*{APPENDIX B: DMFT result}

\begin{figure}
    \centering
    \includegraphics[scale=0.255]{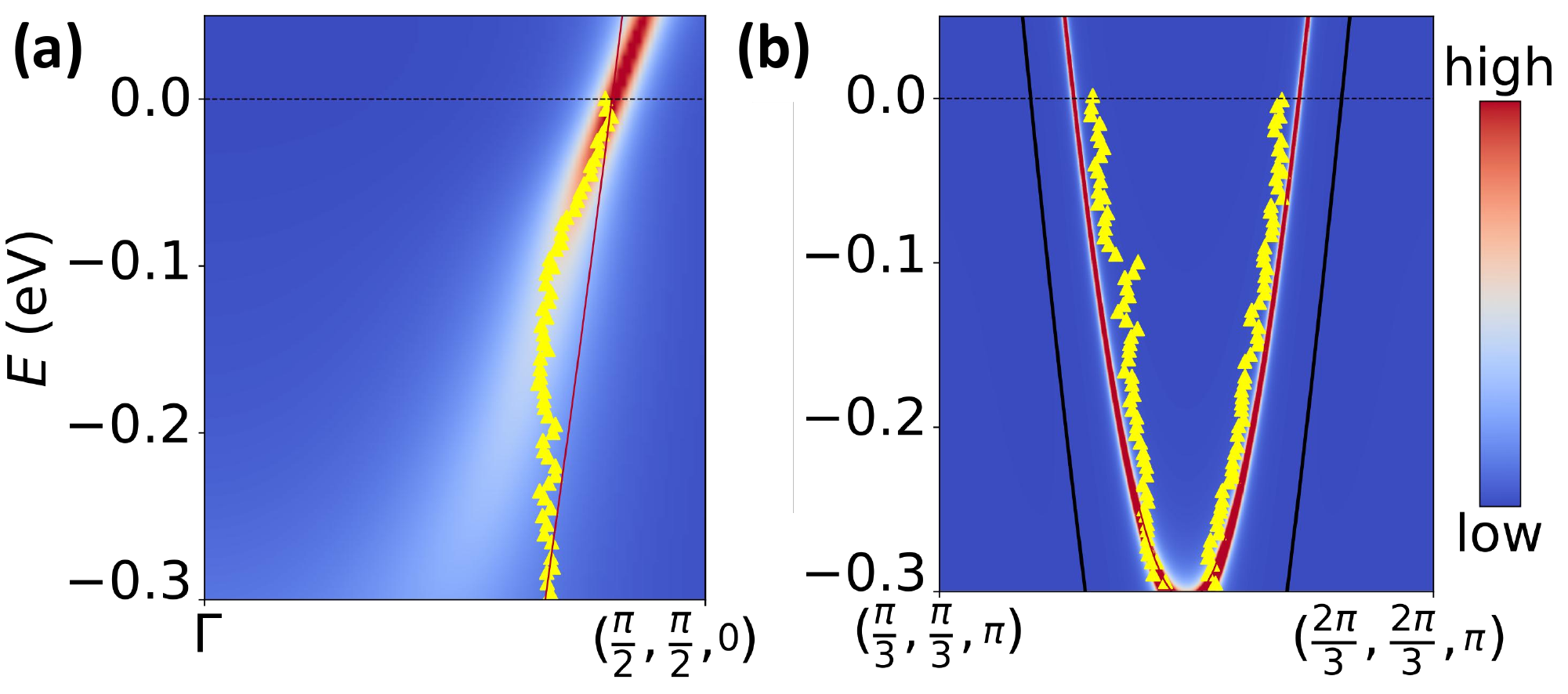}
    \caption{
    Comparison of the DMFT spectral function with experimental data along specific paths. The red solid lines represent the bands obtained from two-band model Hamiltonian, while the yellow triangles denote the experimental data.(a) Spectral function along the path from \(\Gamma\) \((0, 0, 0)\) to \((\pi/2, \pi/2, 0)\).(b) Spectral function along the path from \((\pi/3, \pi/3, \pi)\) to \(\ (2\pi/3, 2\pi/3, \pi)\), the solid black line indicates the DFT result.
    }
    \label{fig:band}
\end{figure}

In Fig.\ref{fig:band}, we provide more comparison between the spectral function results in the \(k_z=0\) plane and the experimental data. Compared to the non-interacting DFT bands, interactions lead to a band renormalization of approximately 2.62. 
Panel (b) of Fig.\ref{fig:band} focuses on the electron-like pocket of 5d electrons near the \(A(\pi, \pi, \pi)\) point in the Brillouin zone. 
We also plot the 5d-band (black line in Fig.\ref{fig:band}(b))from DFT. Although we shift the position of 5d band, the effective mass of 5d band remains almost unchanged between DFT and TB. 
Since the correlation effect is weak for 5d electrons in our approximation, the DMFT calculation does not significantly alter the band structure, and the size and depth of the pocket match well with the experimental data. 
Hence, we can safely say the correlation in 5d is irrelevant to our problem.

\section*{APPENDIX C: DMET result}

\begin{figure}[H]
    \centering
    \includegraphics[scale=0.27]{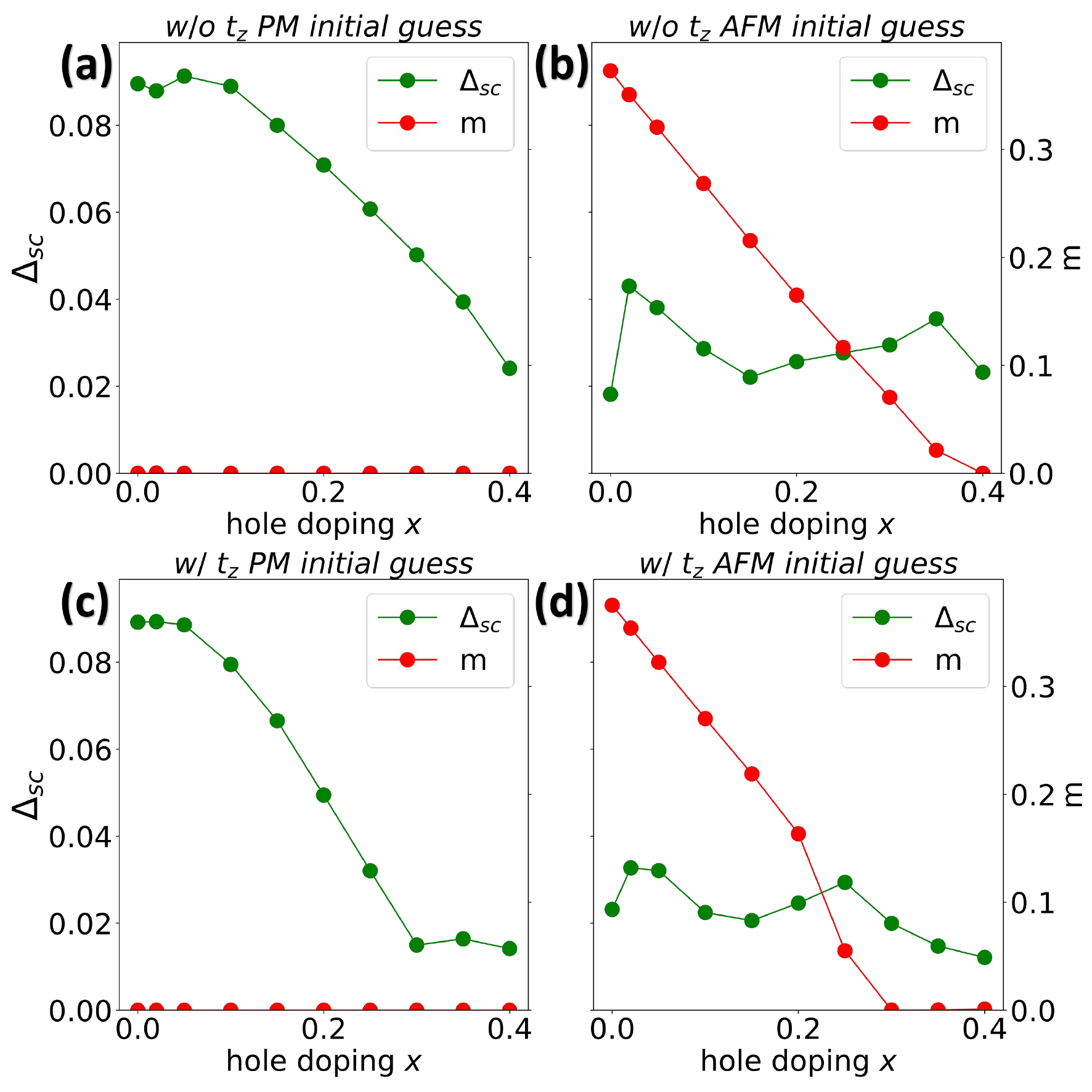}
    \caption{The superconducting and magnetic order parameters as a function of doping, obtained from different initial magnetic states:(a) The model without z-direction hopping and paramagnetic initial state.(b) The model without z-direction hopping and antiferromagnetic initial state.(c) The model with z-direction hopping and paramagnetic initial state.(d) The model with z-direction hopping and antiferromagnetic initial state.
}
    \label{fig:phase}
\end{figure}

The final phase diagrams, both with and without z-direction hopping, are presented in the main text. However, to obtain these phase diagrams, it is crucial to carefully calculate the energy and superconducting pairing symmetry for different initial magnetic states. Fig.\ref{fig:phase} shows order parameter values in different two-band model and different initial magnetic states. For the paramagnetic initial state, whether the model includes z-direction hopping or not, the magnetism is suppressed to very small values. In contrast, for the antiferromagnetic initial guess, the antiferromagnetic order parameter decreases as doping increases, and when the doping reaches around 0.25, even with an antiferromagnetic initial guess, the magnetism converges to a very small value or even to zero.

\begin{figure}[H]
    \centering
    \includegraphics[scale=0.42]{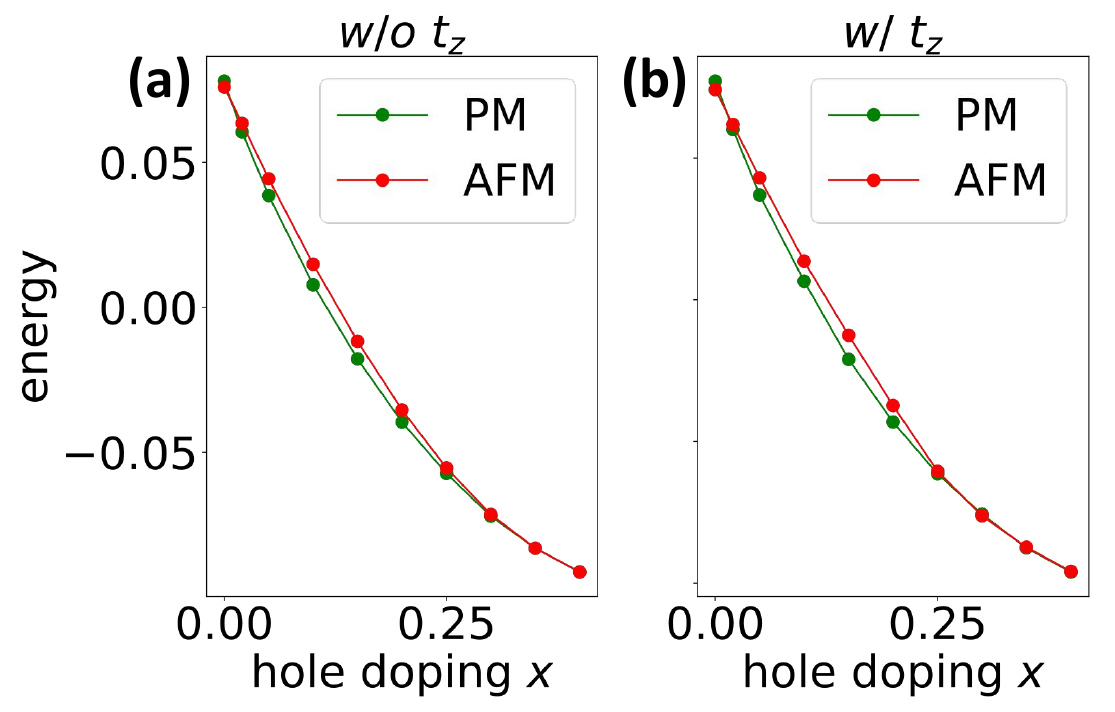}
    \caption{The ground state energies of two-band models under different doping levels, where the green and red lines represent the paramagnetic and antiferromagnetic initial guesses, respectively. 
    (a) Ground state energies obtained by converging from paramagnetic and antiferromagnetic initial guesses under the model parameterswithout z-direction hopping
    Ground state energies obtained by converging from paramagnetic and antiferromagnetic initial guesses under the model parameters with z-direction hopping.}
    
    \label{fig:energy}
\end{figure}

\begin{figure}[H]
    \centering
    \includegraphics[scale=0.41]{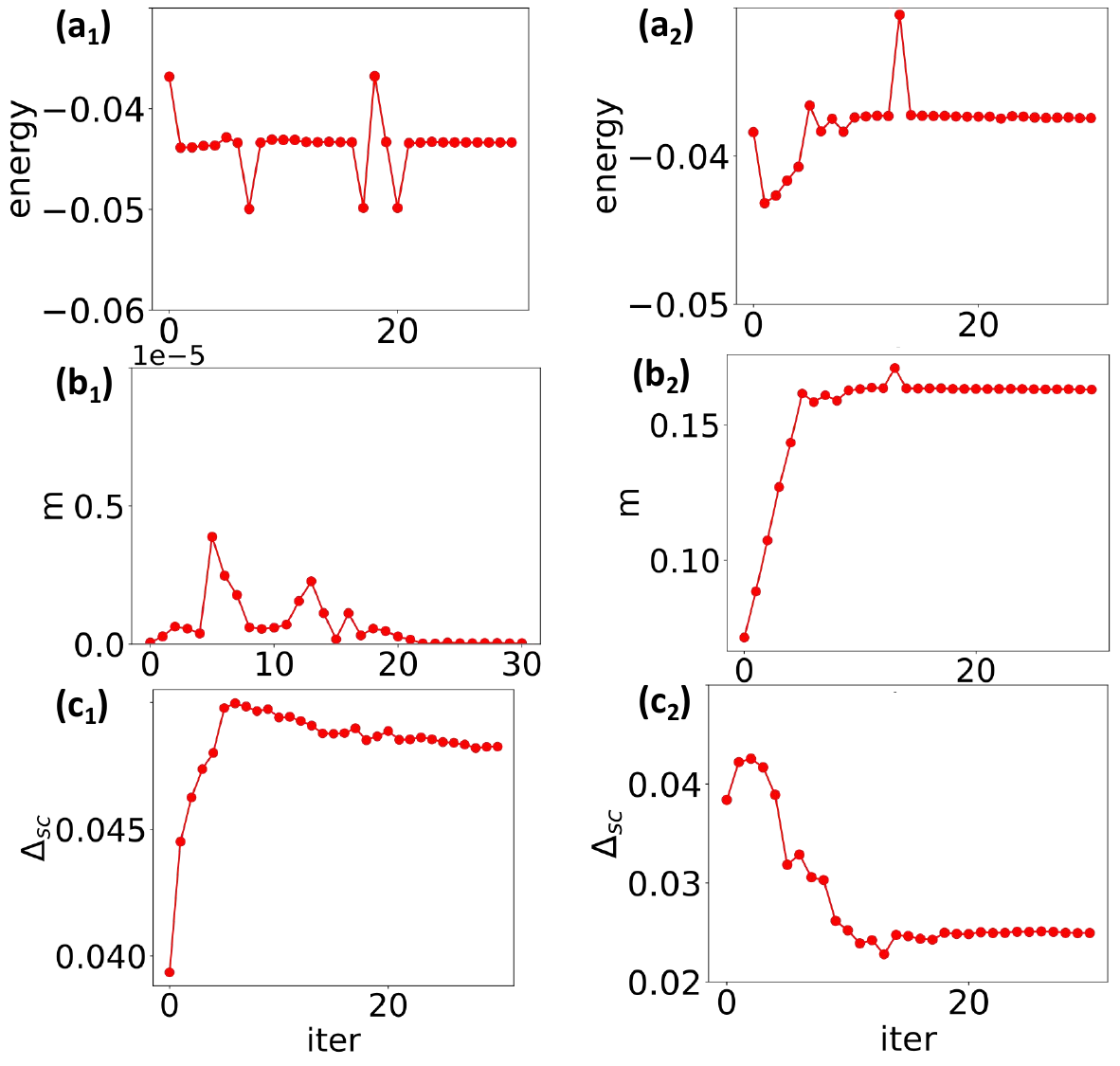}
    \caption{The evolution of energy, magnetism, and superconductivity during the self-consistent iterations. The left column corresponds to the model with z-direction hopping at hole doping $x=0.2$, using the paramagnetic state as the initial guess for the self-consistent iteration, while the right column uses the antiferromagnetic state as the initial guess.(a1), (a2) Energy as a function of the iteration steps.(b1), (b2) Magnetism as a function of the iteration steps.(c1), (c2) Superconducting order parameter as a function of the iteration steps.
}
    \label{fig:n0.8}
\end{figure}

To determine which state corresponds to the true physical state, we need to consider the energy for different initial magnetic states. Fig.\ref{fig:energy} shows the energies after convergence for different initial conditions. It can be observed that when doping is greater than 0.25, the ground state energies for both initial guesses are nearly identical, as can also be seen from the phase diagram in Fig.\ref{fig:phase}. When doping is between 0.02 and 0.25, the paramagnetic ground state energy is always lower than the antiferromagnetic one. It is worth noting that within a very small doping range near half-filling (approximately 0.01), the ground state energy of the antiferromagnetic state becomes lower. This indicates that the system is antiferromagnetic near half-filling, and with the addition of a small amount of doping, the system quickly transitions to a paramagnetic state.

Fig.\ref{fig:n0.8} illustrates the convergence behavior of energy, magnetism, and superconductivity during the DMET calculations. In our computations, the ground state energy and magnetism generally stabilize rapidly after a few iterations. However, the convergence of the superconducting order parameter is slower, with varying speeds depending on the doping range, but it is generally slower than that of energy and magnetism.

\end{document}